\documentstyle[12pt]{article}
\advance \topmargin by -\headheight
\advance \topmargin by -\headsep

\evensidemargin \oddsidemargin
\begin{document}

\topmargin -.6in

\def\rf#1{(\ref{eq:#1})}
\def\lab#1{\label{eq:#1}}
\def\nonu{\nonumber}
\def\br{\begin{eqnarray}}
\def\er{\end{eqnarray}}
\def\be{\begin{equation}}
\def\ee{\end{equation}}
\def\eq{\!\!\!\! &=& \!\!\!\! }
\def\foot#1{\footnotemark\footnotetext{#1}}
\def\lb{\lbrack}
\def\rb{\rbrack}
\def\llangle{\left\langle}
\def\rrangle{\right\rangle}
\def\blangle{\Bigl\langle}
\def\brangle{\Bigr\rangle}
\def\llbrack{\left\lbrack}
\def\rrbrack{\right\rbrack}
\def\lcurl{\left\{}
\def\rcurl{\right\}}
\def\({\left(}
\def\){\right)}
\newcommand\sbr[2]{\left\lbrack\,{#1}\, ,\,{#2}\,\right\rbrack}
\newcommand\pbr[2]{\{\,{#1}\, ,\,{#2}\,\}}
\newcommand\pbbr[2]{\lcurl\,{#1}\, ,\,{#2}\,\rcurl}
\def\v{\vert}
\def\bv{\bigm\vert}
\def\Bgv{\;\Bigg\vert}
\def\bgv{\bigg\vert}
\def\lskip{\vskip\baselineskip\vskip-\parskip\noindent}
\relax

\def\tr{\mathop{\rm tr}}
\def\Tr{\mathop{\rm Tr}}
\newcommand\partder[2]{{{\partial {#1}}\over{\partial {#2}}}}
\newcommand\funcder[2]{{{\delta {#1}}\over{\delta {#2}}}}
\newcommand\Bil[2]{\Bigl\langle {#1} \Bigg\vert {#2} \Bigr\rangle}  
\newcommand\me[2]{\left\langle {#1}\right|\left. {#2} \right\rangle} 

\def\a{\alpha}
\def\b{\beta}
\def\dc{{\cal D}}
\def\d{\delta}
\def\D{\Delta}
\def\eps{\epsilon}
\def\vareps{\varepsilon}
\def\g{\gamma}
\def\G{\Gamma}
\def\grad{\nabla}
\def\h{{1\over 2}}
\def\l{\lambda}
\def\L{\Lambda}
\def\m{\mu}
\def\n{\nu}
\def\o{\over}
\def\om{\omega}
\def\O{\Omega}
\def\p{\phi}
\def\P{\Phi}
\def\pa{\partial}
\def\pr{\prime}
\def\ra{\rightarrow}
\def\s{\sigma}
\def\S{\Sigma}
\def\t{\tau}
\def\th{\theta}
\def\Th{\Theta}
\def\ti{\tilde}
\def\wti{\widetilde}
\newcommand\sumi[1]{\sum_{#1}^{\infty}}   
\def\jc{J^C}
\def\bj{{\bar J}}
\def\sj{{\jmath}}
\def\bsj{{\bar \jmath}}
\def\bp{{\bar \p}}
\def\faa{Fa\'a di Bruno~}
%
\def\lie{{\cal G}}
\def\dlie{{\cal G}^{\ast}}
\def\elie{{\widetilde \lie}}
\def\edlie{{\elie}^{\ast}}
\def\hlie{{\cal H}}
\def\wlie{{\widetilde \lie}}
\def\f#1#2#3 {f^{#1#2}_{#3}}
\def\winf{{\bf w_\infty}}
\def\win1{{\bf w_{1+\infty}}}
\def\Winf{{\bf W_\infty}}
\def\Win1{{\bf W_{1+\infty}}}
\def\hWinf{{\bf {\hat W}_{\infty}}}
\def\Rm#1#2{r(\vec{#1},\vec{#2})}          
\def\OR#1{{\cal O}(R_{#1})}           
\def\ORti{{\cal O}({\widetilde R})}           
\def\AdR#1{Ad_{R_{#1}}}              
\def\dAdR#1{Ad_{R_{#1}^{\ast}}}      
\def\adR#1{ad_{R_{#1}^{\ast}}}       
\def\KP{${\rm \, KP\,}$}                 
\def\KPl{${\rm \,KP}_{\ell}\,$}         
\def\KPo{${\rm \,KP}_{\ell = 0}\,$}         
\def\mKPa{${\rm \,KP}_{\ell = 1}\,$}    
\def\mKPb{${\rm \,KP}_{\ell = 2}\,$}    
%
\def\rlx{\relax\leavevmode}
\def\inbar{\vrule height1.5ex width.4pt depth0pt}
\def\IZ{\rlx\hbox{\sf Z\kern-.4em Z}}
\def\IR{\rlx\hbox{\rm I\kern-.18em R}}
\def\IC{\rlx\hbox{\,$\inbar\kern-.3em{\rm C}$}}
\def\one{\hbox{{1}\kern-.25em\hbox{l}}}
\def\0#1{\relax\ifmmode\mathaccent"7017{#1}%
    	\else\accent23#1\relax\fi}
\def\omz{\0 \omega}
%
\def\ltimes{\mathrel{\vrule height1ex}\joinrel\mathrel\times}
\def\rtimes{\mathrel\times\joinrel\mathrel{\vrule height1ex}}

\def\mark{\noindent{\bf Remark.}\quad}
\def\prop{\noindent{\bf Proposition.}\quad}
\def\theor{\noindent{\bf Theorem.}\quad}
\def\name{\noindent{\bf Definition.}\quad}
\def\exam{\noindent{\bf Example.}\quad}
\def\proof{\noindent{\bf Proof.}\quad}
\newcommand{\nit}{\noindent}
\newcommand{\ct}[1]{\cite{#1}}
\newcommand{\bi}[1]{\bibitem{#1}}
%
%
\def\PRL#1#2#3{{\sl Phys. Rev. Lett.} {\bf#1} (#2) #3}
\def\NPB#1#2#3{{\sl Nucl. Phys.} {\bf B#1} (#2) #3}
\def\NPBFS#1#2#3#4{{\sl Nucl. Phys.} {\bf B#2} [FS#1] (#3) #4}
\def\CMP#1#2#3{{\sl Commun. Math. Phys.} {\bf #1} (#2) #3}
\def\PRD#1#2#3{{\sl Phys. Rev.} {\bf D#1} (#2) #3}
\def\PLA#1#2#3{{\sl Phys. Lett.} {\bf #1A} (#2) #3}
\def\PLB#1#2#3{{\sl Phys. Lett.} {\bf #1B} (#2) #3}
\def\JMP#1#2#3{{\sl J. Math. Phys.} {\bf #1} (#2) #3}
\def\PTP#1#2#3{{\sl Prog. Theor. Phys.} {\bf #1} (#2) #3}
\def\SPTP#1#2#3{{\sl Suppl. Prog. Theor. Phys.} {\bf #1} (#2) #3}
\def\AoP#1#2#3{{\sl Ann. of Phys.} {\bf #1} (#2) #3}
\def\PNAS#1#2#3{{\sl Proc. Natl. Acad. Sci. USA} {\bf #1} (#2) #3}
\def\RMP#1#2#3{{\sl Rev. Mod. Phys.} {\bf #1} (#2) #3}
\def\PR#1#2#3{{\sl Phys. Reports} {\bf #1} (#2) #3}
\def\AoM#1#2#3{{\sl Ann. of Math.} {\bf #1} (#2) #3}
\def\UMN#1#2#3{{\sl Usp. Mat. Nauk} {\bf #1} (#2) #3}
\def\FAP#1#2#3{{\sl Funkt. Anal. Prilozheniya} {\bf #1} (#2) #3}
\def\FAaIA#1#2#3{{\sl Functional Analysis and Its Application} {\bf #1} (#2)
#3}
\def\BAMS#1#2#3{{\sl Bull. Am. Math. Soc.} {\bf #1} (#2) #3}
\def\TAMS#1#2#3{{\sl Trans. Am. Math. Soc.} {\bf #1} (#2) #3}
\def\InvM#1#2#3{{\sl Invent. Math.} {\bf #1} (#2) #3}
\def\LMP#1#2#3{{\sl Letters in Math. Phys.} {\bf #1} (#2) #3}
\def\IJMPA#1#2#3{{\sl Int. J. Mod. Phys.} {\bf A#1} (#2) #3}
\def\AdM#1#2#3{{\sl Advances in Math.} {\bf #1} (#2) #3}
\def\RMaP#1#2#3{{\sl Reports on Math. Phys.} {\bf #1} (#2) #3}
\def\IJM#1#2#3{{\sl Ill. J. Math.} {\bf #1} (#2) #3}
\def\APP#1#2#3{{\sl Acta Phys. Polon.} {\bf #1} (#2) #3}
\def\TMP#1#2#3{{\sl Theor. Mat. Phys.} {\bf #1} (#2) #3}
\def\JPA#1#2#3{{\sl J. Physics} {\bf A#1} (#2) #3}
\def\JSM#1#2#3{{\sl J. Soviet Math.} {\bf #1} (#2) #3}
\def\MPLA#1#2#3{{\sl Mod. Phys. Lett.} {\bf A#1} (#2) #3}
\def\JETP#1#2#3{{\sl Sov. Phys. JETP} {\bf #1} (#2) #3}
\def\JETPL#1#2#3{{\sl  Sov. Phys. JETP Lett.} {\bf #1} (#2) #3}
\def\PHSA#1#2#3{{\sl Physica} {\bf A#1} (#2) #3}
\begin{titlepage}
\vspace*{-1cm}
\noindent
February, 1993 \hfill{IFT-P/011/93}\\
\phantom{bla}
\hfill{UICHEP-TH/93-1} \\
\phantom{bla}
\hfill{hep-th/9302125}
\\
\vskip .3in

\begin{center}
{\large\bf Generalized Miura Transformations, Two-boson KP Hierarchies }
\end{center}
\begin{center}
{\large\bf and Their Reduction to KdV Hierarchies}
\end{center}

\normalsize
\vskip .4in

\begin{center}
{ H. Aratyn\footnotemark
\footnotetext{Work supported in part by U.S. Department of Energy,
contract DE-FG02-84ER40173 and by NSF, grant no. INT-9015799}}

\par \vskip .1in \noindent
Department of Physics \\
University of Illinois at Chicago\\
801 W. Taylor St.\\
Chicago, Illinois 60607-7059\\
\par \vskip .3in

\end{center}

\begin{center}
{L.A. Ferreira\footnotemark
\footnotetext{Work supported in part by CNPq}}, J.F. Gomes$^{\,2}$,
R.T. Medeiros\footnotemark\footnotetext{Supported by CAPES}
and A.H. Zimerman$^{\,2}$

\par \vskip .1in \noindent
Instituto de F\'{\i}sica Te\'{o}rica-UNESP\\
Rua Pamplona 145\\
01405-900 S\~{a}o Paulo, Brazil
\par \vskip .3in

\end{center}

\begin{center}
{\large {\bf ABSTRACT}}\\
\end{center}
\par \vskip .3in \noindent

Bracket preserving gauge equivalence is established between several
two-boson generated KP type of hierarchies.
These KP hierarchies reduce under symplectic reduction (via Dirac constraints)
to KdV, mKdV and Schwarzian KdV hierarchies.
Under this reduction the gauge equivalence is taking form of
the conventional Miura maps between the above KdV type of hierarchies.

\end{titlepage}
\noindent
{\large {\bf 1. Introduction}}
\lskip
The main aim of this paper is to provide arguments for the existence and
usefulness of the linking thread in form of a symplectic
gauge transformation connecting various integrable systems.
It seems to be quite relevant to introduce such an equivalence principle
in view of the growing number of integrable models which recently
entered high energy physics (e.g. matrix models).
The linkage is discussed here in a simple context of two-boson KP
hierarchies. One of advantages of using two-boson systems is that the gauge
map connecting them reduces in the special limit to the well-known Miura
transformation; another is that their provide a common origin for various
KdV type of hierarchies.

We argue that the right setting to study gauge equivalence between integrable
systems is provided by Adler-Kostant-Symes (AKS) \ct{AKS}
theory with a Poisson bracket structure defined in terms of the R-matrix
Lie-Poisson (LP) bracket.
There are several reasons behind this statement.
The theory of classical R-matrices provides a unified approach to
dealing with most, if not all, existing integrable systems.
One practical advantage of using the classical R-matrix theory augmented by
the AKS scheme is that it provides a simple
method to construct commuting integrals for a wide class of integrable models.
In the recent paper \ct{rkp} classical R-matrix theory based on
AKS approach was applied to the algebras of pseudo-differential
symbols. This paper has utilized another feature of AKS approach, namely the
possibility of establishing symplectic gauge invariance between various
related integrable systems (see also \ct{oevelr} for similar approach).
Especially, three integrable models \ct{GKR88,BAK85,rkp}, called
in \ct{rkp} as ${\rm KP}_{\ell}$ with $\ell$ taking values $0,1$ and $2$
were shown to be connected by the symplectic gauge transformation, acting on
Lax operators parametrizing the coadjoint orbits.
Symplectic character of the gauge mapping
ensured that the R-matrix LP bracket was preserved.
Note, that for $\ell =0$ the AKS system ${\rm KP}_{0}$ is nothing but a
standard KP hierarchy.

In Section 2 we briefly recapitulate the AKS method as applied to the algebras
of pseudo-differential symbols and state the results of \ct{rkp} concerning
the gauge equivalence of relevant integrable models.
In Section 3 we explain the status of the two-boson KP hierarchy, which
appears in this setting as an invariant subspace of the coadjoint orbit
within the \mKPa hierarchy.
We will work with two main cases of two-boson KP hierarchies, one defined
within \mKPa hierarchy will be
called \faa KP hierarchy, while the second defined
within ${\rm KP}$ hierarchy for a quadratic two-boson KP
hierarchy.
We will establish for them the gauge invariance playing the role of
generalized Miura transformations.
We emphasize the symplectic character of equivalence of \mKPa and
${\rm KP}$ and show how this feature explains the 2-boson representation of
$\Win1$ and $\hWinf$ in terms of the \faa polynomials.
We also made a point that the gauge equivalence established for two-boson
systems is valid for an arbitrary n-th Poisson bracket structure
and not only the first Poisson bracket structure.
In Section 4 we apply Dirac reduction scheme to both two-boson KP
hierarchies. We obtain in the process of reduction the standard
``one-boson" KdV and mKdV hierachies. On reduced manifold the gauge
transformation connecting the two models takes the form of the Miura
transformation.
We also present some comments on generalizing Schwarzian KdV
(SKdV) hierarchy to the two-boson system.

The generalized Miura maps appeared before in literature
(see \ct{BAK85,fordy,schiff}) without however being
studied in terms of the gauge transformations between Lax operators and
connected with conventional Miura maps via Dirac reduction.
\lskip
{\large {\bf 2. AKS Construction of Generalized KP Hierarchies}}
\lskip
Here we will apply the AKS construction \ct{AKS} on Lie algebra $\lie $
of pseudo-differential operators on a circle. An element
of $\lie$ is an arbitrary
pseudo-differential operator $X = \sum_{k \geq -\infty}
D^k \,X_k (x) \,$.
An infinitesimal version of adjoint transformation is given
by $ad (Y) X= \sbr{Y}{X}$.
In this setting, an identification of the dual space $\dlie$ with $\lie$ and
of the coadjoint action with the adjoint action is allowed by the Adler trace;
an invariant, non-degenerate
bilinear form given by $\me{L}{X} \equiv {\Tr}_A \( L X \) =
\int {\rm Res} L X$.

There exist three decompositions of $\lie$ into a linear sum of two
subalgebras \ct{GKR88,BAK85,rkp} (i.e. $\lie = \lie_{+}^{\ell} \oplus
\lie_{-}^{\ell}$, with index $\ell$ taking values $\ell = 0,1,2$):
\be
\lie^{\ell}_{+} = \{\, X_{\geq \ell}
= \sumi{i=\ell} D^i X_i (x) \,\}
\quad;\quad
\lie^{\ell}_{-} = \{\, X_{< \ell}
= \sumi{i=-\ell+1 } D^{-i} X_{-i}(x)\,\}
\lab{subalg}
\ee

The dual spaces to subalgebras $\lie^{\ell}_{\pm}$ are given
by:
\be
{\lie^{\ell}_{+}}^{\ast} = \{ L_{< -\ell}
= \sumi{i=\ell+1} u_{-i}(x) D^{-i}\, \} \;\; ;\;\;
{\lie^{\ell}_{-}}^{\ast} = \{  L_{\geq -\ell} =
\sumi{i=-\ell} u_{i}(x) D^{i}\}
\lab{dsubalg}
\ee

All three decompositions give rise to integrable models via the
AKS construction. Define the R-matrix for all the above cases as
$R_{\ell} \equiv P_{+}^{\ell}  - P_{-}^{\ell} $, where
$P_{\pm}^{\ell}$ are projections on $\lie_{\pm}^{\ell}$.
It follows from the general formalism that $\lb X , Y \rb_{R_{\ell}}
\equiv \lb R X , Y \rb/2 + \lb X, R Y \rb /2
=\lb X_{\geq \ell},
Y_{\geq \ell}\rb - \lb X_{< \ell} , Y_{< \ell} \rb$
defines an additional (with respect to usual commutator)
Lie structure on $\lie\,$ (see \ct{rkp} and references therein).
This additional structure gives rise to the LP $R$-bracket
for $F, H \in C^{\infty} \(\dlie, \IR \)$:
\be
\{ F \, , \, H \}_R (L) =
\me{L}{{\sbr{\nabla F (L)}{\nabla H (L)}}_R}   \lab{Rbra}
\ee
where the gradient $\nabla F\,: \, \dlie \ra \lie$ is defined by the
standard formula  given in \ct{AKS,rkp}.
The AKS scheme defines the Hamiltonian equations of motion
to be $d F / dt = \{ H , F \}_{R}$ for
$F \in C^{\infty} \(\dlie, \IR \)$.
The basic result of AKS formalism states that the $Ad^{\ast}$-invariant
functions (Casimirs) Poisson commute on $\( \dlie, \{ \cdot ,\cdot\}_R \)$
establishing integrability of the system and providing quantities in
involution.

{}From the general relation for the $R$-coadjoint action of
$\lie\,$ on its dual space we find
that the infinitesimal shift along an $R$-coadjoint orbit
$\OR{\ell}\,$ has the form: $\d_{R_\ell} L \equiv
ad^{\ast}_{R_\ell} (X) L =
\lb X_{\geq \ell} , L_{< -\ell} \rb_{< -\ell}
- \lb X_{< \ell} , L_{\geq -\ell} \rb_{\geq -\ell}$.

We will now focus on the Hamiltonian structure of the
integrable systems given by decompositions labelled by $\ell=0,1$
(as in \ct{rkp} we call them here as \KPl hierarchies).
For the remaining $\ell=2$ model we refer the
reader to \ct{rkp}.
\lskip
{\sl \KPl Hierarchies.}
The \KPo model is defined on the manifold being the $R$-coadjoint orbit of the
form $\, \OR{0} = \Bigl\{ L =  D+ \sumi{k=1}
\, u_k (x) \, D^{-k} \Bigr\} \,$.
The functions $H_{r+1} = { 1 \o {r+1}} \int {\rm Res}
L^{r+1}\,$ are Casimir functions on $\dlie$,
which in $R$-matrix approach \ct{AKS} produce commuting integrals of motion.
In fact we find for this model
\be
\pa L/ \pa t_r = \h ad^{\ast} \Bigl( (\nabla H_{r+1})_{+} -
(\nabla H_{r+1})_{-} \Bigr)\, L
= \sbr{\( (L^r )_{+} \)}{L}
\lab{rkp0}
\ee
The subscript $(+)$ means taking the purely differential part of $\,
L^r \,$ and $t = \{ t_r \} $ are the evolution parameters
(infinitely many time coordinates).
We recognize in \rf{rkp0} the standard $KP$ flow equation.
The flows \rf{rkp0} are bi-Hamiltonian \ct{dickey},
i.e. there exist two
Poisson bracket structures $\{\cdot , \cdot \}_{1,2}$ , such that
we can rewrite \rf{rkp0} as a Lenard recursion relation
$
\pa {L} / \pa {t_r} = \{ H_r \,, \, L \}_2= \{ H_{r+1} \,, \, L \}_1
$
for the hierarchy of Poisson bracket structures with $r=1,2, \ldots$.
The first Hamiltonian structure is found to be induced by the
LP structure: $ \{ u_i (x) \, , \, u_j (y) \}_{R_0} = \O^{(\ell=0)}_{i-1,j-1}
(u(x))\,\d (x-y)$, with \ct{watanabe,BAK85}:
\be
\O^{(\ell)}_{i,j}\(u(x)\) = - \sum_{k=0}^{i+\ell}
{i+\ell\choose k} u_{i+j+\ell - k+1} (x) D^k_x +
\sum_{k=0}^{j+\ell} (-1)^k{j+\ell\choose k} D^k_x u_{i+j+\ell -k+1} (x)
\lab{watform}
\ee
This LP bracket algebra is isomorphic to
the centreless
$\Win1\,$ algebra \ct{wu91}. All this just classifies \KPo as
the standard $KP$ hierarchy.

We now turn our attention to \mKPa hierarchy. Here the Lax operator takes
the form
$ L^{(1)} = D + u_0 +
\, u_1 D^{-1} + \sum_{i \geq 2} \, v_{i-2} D^{-i} \,$.
Application of \rf{Rbra} gives a Hamiltonian structure that is a
direct sum of the $2\times 2$ matrix $P^{(1)}$ with non-zero matrix elements
$P^{(1)}_{12} = P^{(1)}_{21} = D$ associated with the modes
$\{ u_0, u_1 \}$ and the Hamiltonian structure $\O^{(1)}$ from \rf{watform}
associated  with $\{ v_i \v i \geq 0\}$ \ct{BAK85}.
Note that $\O^{(1)}\,$ corresponds to the centreless $\Winf\,$ algebra.
\lskip
{\sl ``Gauge" Equivalence of \mKPa Hierarchy to
Ordinary \KP.}
The fundamental result of \ct{rkp} was the proof
that all three hierarchies are ``gauge" equivalent via
 generalized Miura transformations.
Here we focus on the link between two \KPl systems discussed above.
Reference \ct{rkp} presented a symplectic (Hamiltonian)
map between the orbits such that $\, G: {\cal O}(R_1)
\longrightarrow {\cal O}(R_0)\,$.
The term ``symplectic" (``Hamiltonian") means that under the
map $G\,$, the LP bracket structure on ${\cal O}(R_1)\,$
is mapped into the LP bracket structure on ${\cal O}(R_0)$.
\be
\left\{ F_1 , F_2 \right\}_{R_0}
\( G (L)\)
= \left\{ F_1 \( G (L)\)\, ,\, F_2 \( G (L)\)\right\}_{R_1}
\lab{3-2}
\ee
where $F_{1,2}\,$ are arbitrary functions on ${\cal O}(R_0)$ and
$L\,$ and $G (L)\,$ denote
coordinates on the orbits ${\cal O}(R_1)$ and ${\cal O}(R_0)$, respectively.
As a consequence of \rf{3-2}, the infinite set of involutive integrals
of motion $\{ {\wti H}_N \lb G (L) \rb \}\,$ of the integrable system
on ${\cal O}(R_0)$ are transformed into those of the integrable system on
${\cal O}(R_1)$: $H_N \lb L\rb = {\wti H}_N \lb G(L) \rb\,$ .

As shown in \ct{rkp} the right choice of the
map $\, G: {\cal O}(R_1)\longrightarrow {\cal O}(R_0)\;$ is given by
\be
G (L^{(1)})= D + \sumi{k=1} {\wti u}_k (x)\, D^{-k}
 = Ad^{\ast} ( g (L^{(1)}) ) ( D + u_0 (x) + u_1 (x) D^{-1} +
\sumi{k=2}  v_{k-2} (x) D^{-k})
\lab{3-3}
\ee
where the group element $ g(L^{(1)}) $  depends on $L^{(1)}$ in
${\cal O}(R_1) \subset \dlie \,$.

It is in the sense of eqs. \rf{3-2} and \rf{3-3} that
the integrable systems on the orbits $\OR{}$ for different $R$-matrices
are called ``gauge" equivalent.
We will provide further arguments in Section 4 for that the mapping of one
Poisson bracket structure of an integrable model into another one by the
group coadjoint action \rf{3-3} deserves a name of the
generalized Miura transformation.

Due to the simple formula
$\exp (\p_0 (x)) D \exp (- \p_0 (x)) = D - \pa_x \p_0 (x)$,
it is easy to see that a ``gauge" generator in \rf{3-3} must be given by
\be
g (L) = \exp \p_0 (x) \;\;\;\;  ,\;\;\;\;   \pa_x \p_0 (x) = u_0 (x)
\lab{3-5}
\ee
The proof for gauge equivalence in \ct{rkp} amounted to verifying
\rf{3-3} with \rf{3-5}.
This established the ``gauge" equivalence of \mKPa and \KP
by explicitly constructing the generalized Miura-like transformation
\rf{3-3}-\rf{3-5}, which maps the Poisson bracket structure of \KP into that
of \mKPa and vice versa.
\lskip
{\large {\bf 3. Main Two Two-Boson KP Systems}}
\lskip
{\sl \faa Hierarchy.}
Here we go back to \mKPa and make the following crucial observation.
Consider truncated elements of
${{\cal G}_{-}^1}^{\ast} $ of the type $L_{J}^{(1)} = D + u_0 +  \,u_1\,
D^{-1} = D - J + \bj D^{-1} $, where we have introduced two Bose currents
$(J,\bj)$ to create fit with notation used in \ct{2boson}.
One easily verifies that under the coadjoint action $\d_{R_1} L_{J}^{(1)} =
ad^{\ast}_{R_1} (X) L_{J}^{(1)}$ this finite Lax operator maintains its form,
i.e. the two-boson Lax operators span an $R_1$-orbit of finite functional
dimension $\, 2$.
This observation, already present in \ct{GKR88} clarifies status of
two-boson $(J,\bj)$ system as a consistent restriction of the full \mKPa
hierarchy understood as an orbit model.
Note that there are only two possibilities for the invariant $R_1$-orbit;
the two-boson system and the full \mKPa system (in quasiclassical limit
situation is much richer).
A calculation of the Poisson bracket according to \rf{Rbra}:
$\{ \me{L_{J}^{(1)}}{X} \, ,\,  \me{L_{J}^{(1)}}{Y} \}_{R_1} =
\me{L_{J}^{(1)}}{\lb X\, , \, Y \rb_{R_1} }$
yields the first bracket structure  of two-boson $(J,\bj)$ system given as
LP $R$-bracket: $\pbr{J(x)}{\bj(y)}= - \d^{\pr} (x-y)$ and zero
otherwise.

As we show now the two-boson KP hierarchy is associated with so called
\faa polynomials and we will call it
therefore \faa hierarchy.
Consider namely the gauge transformation between \mKPa and \KPo
generated by $\P$ such that $\P^{\pr} = J$:
\be
L_{J} = e^{-\P} L_{J}^{(1)} e^{\P} = D + \bj \(D + J\)^{-1} =
D + \sumi{n=0} (-1)^n \bj P_n (J) D^{-1-n}
 \lab{faalax}
\ee
where $P_n (J)= (D + J)^n \cdot 1$ are the \faa polynomials.
As a corollary of the symplectic character of the ``gauge"
transformation used in \rf{faalax}, we conclude that
$u_n = (-1)^n \bj P_n (J) $ satisfy the
Poisson-bracket $\Win1\,$ algebra described by the form $\O^{(0)}$
{}from \rf{watform} \ct{BAK85,2boson}.
It is possible to introduce a deformation parameter into
the \faa representation of $\Win1\,$ algebra by redefining $u_n$ to
$u_n (h) = (-1)^n \bj (h D + J)^n \cdot 1$ \ct{2boson}.
Now the semiclassical limit is simply obtained by taking $h \to 0$
in $u_n (h)$ and yields the generators of ${\bf w_{1+\infty}}$
algebra.

The higher bracket structures have been investigated in \ct{BAK85,2boson}
and the result can be summarized as follows.
The three lowest Hamiltonian functions are:
\be
H_{J\,1} = \int \bj \quad;\quad H_{J \,2} = \int - \bj J
\quad;\quad
H_{J\,3} = \int \( \bj J^2 + \bj J^{\pr} + \bj^2 \) \lab{hJ}
\ee
For the general Hamiltonian matrix structure $P_i$ we have
\be
\partder{}{t_r} { J \choose \bj} = P_{J\,i} \,
{ {\d H_{J\, r+2-i}}/ {\d J} \choose
{\d H_{J\,r+2-i}}/ {\d \bj}} = P_{J\,1} {\d H_{r+1}/\d J\choose
\d H_{r+1}/ \d \bj } = P_{J\,2} { \d H_{r}/ \d J \choose
\d H_{r}/ \d \bj}
\lab{iflow}
\ee
Among the multi-Hamiltonian structures only $P_{J\,1}$ and $P_{J\,2}$ are
independent.
All other matrices $P_{J\,i}\;,\; i=3,4,\ldots$ are related
to $P_{J\,2}$ through $P_{J\,i} =\(P_{J\,2} (P_{J\,1})^{-1} \)^{i-2} P_{J\,2}$
involving the so-called recurrence matrix $P_{J\,2} (P_{J\,1})^{-1} $
\ct{fordy,2boson}.
The explicit form of first and second local Hamiltonian structures is:
\be
P_{J\,1} = \left(\begin{array}{cc}
0 & - D \\
-D & \; 0 \end{array}
\right) \;\; , \;\;
P_{J\, 2} =\left(\begin{array}{cc}
2 D & \; D^2 + D J \\
- D^2 + J D &\; D \bj+ \bj D\end{array}
\right)
\lab{Jp1Jp2}
\ee
Taking $r=2$ in \rf{iflow} we especially get the Boussinesq equation:
\be
J_{t_2} = {\pbr{J}{H_{J\,3}}}_1 = -h J^{\pr \pr} - \( J^2 \)^{\pr} - 2
\bj^{\pr}  \;\; ; \; \;
\bj_{t_2} ={\pbr{\bj}{H_{J\,3}}}_1 = h \bj^{\pr \pr} - 2 \(\bj J\)^{\pr}
\lab{boussin}
\ee
where we re-introduced $h$ as a deformation parameter.
In the dispersiveless limit $h \to 0$ taken in \rf{boussin}
we obtain the classical dispersiveless long wave equations (Benney
equations) \ct{LM79,BAK85}.
\lskip
{\sl Quadratic Two-Boson KP Hierarchy.}
Here we call quadratic two-boson KP hierarchy
the construction presented by Wu and Yu \ct{wuyu}
in order to realize ${\hat W}_{\infty}$ as a hidden current algebra in the
2d $SL(2, \IR )/U(1)$ coset model.
Construction is based on the pseudo-differential operator:
\be
L_{\sj} = D + \bsj\, \(D - \sj \,- \bsj \,\)^{-1} \sj
\lab{sjlax}
\ee
Let us discuss the Hamiltonian structure first.
The three lowest Hamiltonian functions are:
\be
H_{\sj\;1} \! = \! \! \int \! \sj\, \bsj \; ;\;
H_{\sj \;2} \! = \! \! \int \! - \sj^{\pr}\, \bsj + \sj^{2}\,
\bsj +\sj\, \bsj^{2}  \; ;\;
H_{\sj\;3} \! = \! \! \int \! \sj^{\pr \pr}\, \bsj -
3 \sj \,\sj^{\pr}\, \bsj
- 2 \sj^{\pr}\, \bsj^2 - \sj\, \bsj\, \bsj^{\,\pr} + \sj^{3} \, \bsj
+ 3 \sj^{2} \, \bsj^{\,2} + \sj\, \bsj^{\,3}
 \lab{hsj3}
\ee
Among the Hamiltonian structures only second and third are local and are
given by
\be
P_{\sj\;2}= \left(\begin{array}{cc}
0 & D \\
D & \; 0 \end{array}
\right) \;\; , \;\;
P_{\sj \;3}=\left(\begin{array}{cc}
D \sj \,+ \sj\, D & \; -D^2 + D \sj\, + \bsj \, D \\
D^2 + \sj\, D+ D \bsj \, &\; D\bsj\, + \bsj \, D \end{array}
\right)
\lab{p1sjp2sj}
\ee
\prop  The Hamiltonian structure corresponding to the Lax operator
$L_{\sj}$ in \rf{sjlax} is invariant under the following
two transformations:
\br
\sj & \to & \bsj - {\sj^{\pr} \o \sj} \qquad \quad \quad  \mbox{\rm and}
 \qquad \quad \quad \bsj \to \sj
\lab{sjtransf} \\
\bsj &\to & \sj + { \bsj^{\,\pr} \o \bsj \, }\qquad \quad\quad \mbox{\rm and}
   \qquad \quad \quad \sj \to \bsj
\lab{sjtransf2}
\er
\proof One verifies relatively easily that the bracket structures induced by
 both $P_{\sj\;2}$ and
$P_{\sj\;3}$ are invariant under the transformations \rf{sjtransf} and
\rf{sjtransf2}.
Since $P_1= P_2 P_3^{-1} P_2$, a recurrence matrix $P_2 (P_1)^{-1}$ and
all remaining higher hamiltonian structures must therefore remain invariant
under \rf{sjtransf} and \rf{sjtransf2}.
This completes the proof. One can also directly verify that
all three Hamiltonians \rf{hsj3} are invariant under
\rf{sjtransf} and \rf{sjtransf2}.
Hence we conclude that the Lax operators given by
\br
L_{\sj} \eq D + \sj\, \(D - \sj \,- \bsj \,+ {\sj^{\,\pr} \o \sj} \)^{-1}
\( \bsj - {\sj^{\pr} \o \sj} \)
\lab{sjlax1} \\
L_{\sj} \eq D + \( \sj + {\bsj^{\, \pr} \o \bsj} \)
\(D - \sj \,- \bsj \,-{\bsj^{\,\pr} \o \bsj} \)^{-1}  \, \bsj
\lab{sjlax2}
\er
lead to the same Hamiltonian functions as \rf{sjlax}.
\lskip
{\sl Gauge Equivalence between \faa and Quadratic Two-Boson Hierarchies.
Generalized Miura Map.}
We apply on $L_{\sj}$ from \rf{sjlax} the gauge transformation
generated by $\xi  = \( \p +\bp - \ln \sj\,\)$ with result:
\be
L_{\sj} \to \exp ( - \xi )\,L_{\sj}\,
\exp (\xi ) = D + \sj \,+\bsj \, +
\sj \, ( \sj^{-1} \, )^{\pr} \, +  \bsj \, \sj \,
D^{-1} = D - J + \bj D^{-1}   \lab{lsjgauge}
\lab{gaugea}
\ee
where we have introduced
\be
J = - \sj \,- \bsj \, + {\sj^{\,\pr}  \o \sj}  \qquad ; \qquad
\bj = \bsj \, \sj            \lab{gmiura}
\ee
One can now verify explicitly that with the bracket structure given by
$P_{\sj\;2}$ in \rf{p1sjp2sj} variables defined in \rf{gmiura} satisfy
the second bracket structure $P_{J\, 2}$ \rf{Jp1Jp2} of \faa hierarchy.
As a corollary we obtained therefore a short proof for the quadratic
two-boson KP hierarchy \ct{wuyu} system realizing $\hWinf$.
We also obtained a Miura transform for two-Bose hierarchies in
form of \rf{gmiura} which generalizes the usual Miura transformation
between one-bose KdV and mKdV structures (as given below).

It is intriguing that the higher hamiltonian structures of quadratic
two-boson hierarchy are being mapped by \rf{gmiura} to their counterparts
in \faa hierarchy while the gauge equivalence established in Section 1 is
limited to the first bracket structure.
Explanation for the equivalence of higher structures follows however easily
{}from two additional features.
First, it is true that the Hamiltonian functions are invariant under the gauge
equivalence. Second, we note that the Lenard recursion relations extend to
two-boson system in \mKPa as observed in \ct{rkp}.
One can now use the above two facts to extend the gauge equivalence (in the
symplectic sense) to the arbitrary order of bracket structure
for the two-boson systems.

Let us go back to the alternative expression \rf{sjlax1} for the quadratic
two-boson hierarchy. It can be rewritten under multiplication by $1=\sj\,
\sj^{\,-1}$ from the right and left as $
L_{\sj} = 1\, L_{\sj}\, 1 = \sj\, \sj^{\,-1}\,L_{\sj}\, \sj\, \sj^{\,-1}\,
= D + \, \(D - \sj \,- \bsj \,\)^{-1} \( \bsj\, \sj \,- \sj^{\,\pr} \, \)
$.
Next step is to gauge transform $L_{\sj}$ from KP to ${\rm KP}_1$ hierarchy
by acting with gauge transformation generated by $\exp ( \p + \bp )$ obtaining
\be
L_{\sj} \sim \exp \(- \p - \bp \) \,L_{\sj} \exp \( \p + \bp \)
= D + \sj \,+\bsj \,+\, D^{-1} \( \bsj\, \sj \,- \sj^{\,\pr} \, \)
\lab{sjlaxb}
\ee
which is of the form of the \faa hierarchy (up to conjugation) with
$ J =- \sj \,-\bsj \,$ and $ \bj = \bsj\, \sj \,- \sj^{\,\pr} \, $.
This reproduces construction given in \ct{2boson} to get the second bracket
structure from the first.
Note that under \rf{sjtransf2} this is transformed into
$ J = -\sj \,-\,\bsj \, - {\bsj{\,\pr} \o \bsj }\,$ and
$ \bj = \bsj\, \sj \,$ differing from \rf{gmiura} by a conjugation
$\sj\, \leftrightarrow \bsj \,$.

Similarly for \rf{sjlax2} we find $L_{\sj} = \bsj^{\,-1} \bsj \,
L_{\sj} \bsj^{\,-1} \bsj \,=\,
D + \(\bsj\, \sj + \bsj^{\, \pr} \,\)
\(D - \sj \,- \bsj \,\)^{-1}
$.
The same transformation as in \rf{sjlaxb} gives
\be
L_{\sj} \sim \exp \(- \p - \bp \) \,L_{\sj} \exp \( \p + \bp \)
= D + \sj \,+\bsj \,+\, \(\bsj\, \sj + \bsj^{\, \pr} \,\) D^{-1}
\lab{sjlaxd}
\ee
producing ${\rm KP}_1$ object with
$ J =- \sj \,-\bsj \,$ and $ \bj = \bsj\, \sj \,+\bsj^{\,\pr} \, $.
This time under \rf{sjtransf} these variables are transformed into
$ J = -\sj \,-\,\bsj \,+ {\sj{\,\pr} \o \sj }\,$ and
$ \bj = \bsj\, \sj \,$ identical to \rf{gmiura}.

We see that because of \rf{sjtransf} and \rf{sjtransf2} there is an ambiguity
in the possible form of generalized Miura transformation and \rf{gmiura} can
appear also in other forms. All of them are connecting the Poisson bracket
structure of \faa hierarchy with the corresponding Poisson bracket structure
of the quadratic two-boson hierarchy.
\lskip
{\sl AKNS/NLS Hierarchy.}
AKNS or NLS system is a constrained KP system described by:
\be
L_{AKNS}\, =\, D + {\bar \Psi} D^{-1} \Psi
\lab{akns}
\ee
with the first two bracket structures given by (see for instance \ct{oevels}):
\be
P_{AKNS\,1}= \left(\begin{array}{cc}
0 & 1 \\
-1& \; 0 \end{array}
\right) \;\; , \;\;
P_{AKNS\,2}=\left(\begin{array}{cc}
-2 {\bar \Psi} D^{-1} {\bar \Psi} & \; D + 2 {\bar \Psi} D^{-1} \Psi\\
D + 2 \Psi D^{-1} {\bar \Psi}\, &\; - 2 \Psi D^{-1} \Psi \end{array}
\right)
\lab{aknsp1p2}
\ee
One can easily show that AKNS hierarchy is equivalent to \faa two-boson KP.
The proof is based, in the spirit of \ct{rkp}, on
establishing gauge transformation between two hierarchies.
We show now the argument to illustrate the power of gauge
transformation argument in the KP setting.
Consider $ L_{AKNS} \to G^{-1} L_{AKNS} G = G^{-1} D G +
G^{-1} {\bar \Psi} \, D^{-1} \Psi G$.
Choose $ G^{-1} = \Psi$, which leads to
\be
G^{-1} L_{AKNS} G = \Psi \, D \Psi^{-1} \, +  {\bar \Psi} \, \Psi \, D^{-1}
= D + \Psi \, ( \Psi^{-1} \, )^{\pr} \, +  {\bar \Psi}\, \Psi \, D^{-1}
\lab{akns4}
\ee
Clearly the gauge transformed $L_{AKNS}$ is an element of ${\rm KP}_1$
hierarchy and it is therefore natural to introduce new variables such
that $ J = - \Psi \, ( \Psi^{-1} \, )^{\pr} $ and $\bj = {\bar \Psi}\,
\Psi$ and the inverse relation being $\Psi = \exp \(\int J\)$ and
${\bar \Psi} = \bj \exp -\(\int J\)$.
Since we now have established a gauge equivalence between two hierarchies
it is clear that the first bracket structure in \rf{aknsp1p2}
leads to $\{ \bj (x) \, , \, J (y) \}_1 = - \d^{\pr} (x-y)$ and
therefore a linear $\Win1$ algebra.
The second bracket structure in \rf{aknsp1p2} leads to the second structure
in \rf{Jp1Jp2} and correspondingly non-linear $\hWinf$.
If we only took the linear structure in $P_{AKNS\,2}$ (i.e.
$\pbr{{\bar \Psi}\, (x)}{\Psi\,(y)} = \d^{\pr} (x-y)$) we would have induced
\rf{Jp1Jp2} in its ``un-deformed" form  with upper left corner of
$P_{J \, 2}$ in \rf{Jp1Jp2} being zero, corresponding to $\O^{(1)}$ or
$\Winf$.

The AKNS system is also gauge equivalent to quadratic KP hierarchy if
we make in \rf{akns4} a substitution ${\bar \Psi} = \bsj\, \exp (\phi +\bp)$
and $ \Psi = \exp (-\phi- \bp) \sj$ or inversely
$\Psi^{\pr}/\Psi = - \sj\, - \bsj \, + \sj^{\, \pr}/\sj$
and ${\bar \Psi} \Psi = \bsj\, \sj$.
\lskip
{\large {\bf 4. Reduction to ``One-boson" KdV Systems}}
\lskip
We apply here the Dirac reduction scheme to obtain one-boson hierarchies
{}from two-boson hierarchies. The general feature will be a transformation
of some two-boson Hamiltonian equations of motion expressed by 2-nd bracket
structure $\d \G/ \d {t_r} = \{ \G \,, \, H_r \}_2$
(where $\G$ denote original degrees of freedom)
to one-boson Hamiltonian system according to the Dirac scheme:
\be
\partder{X}{t_r} = \{ X \,, \, H^{D}_r \}_{Dirac}
\lab{diraham}
\ee
with $X$ denoting a surviving one-boson degree of freedom.
Another point is that it is a presence of symmetry in \rf{sjtransf}
and \rf{sjtransf2} that enables reduction to be made.
\lskip
{\sl KdV Hierarchy.}
Consider the Dirac constraint: $\Theta = J=0$ for system in \rf{faalax}.
First let us discuss the resulting Dirac bracket structure.
We find for the surviving variable $\bj$:
\br
\{ \bj (x) \, , \, \bj (y) \}_2^{D} \eq \{ \bj (x) \, , \, \bj (y) \}_2
- \int dz dz^{\pr} \{ \bj (x) \, , \, \Theta (z) \}_2 \{ \Theta (z) , \Theta
 (z^{\pr} )
\}_2^{-1} \{ \Theta (z^{\pr}) \, , \, \bj (y) \}_2  \nonu \\
\eq 2 \bj (x) \d^{\pr} (x-y) +\bj^{\pr} (x) \d (x-y) +
\h \d^{\pr \pr \pr} (x-y)
\lab{dira2}
\er
The reduced Lax operator looks now as:
\be
l_{J} = D + \bj D^{-1}  \lab{faa3}
\ee
and the corresponding (non-zero) lowest Hamiltonian functions $H^{KdV}_r
\equiv \Tr l_{J}^r /r$ are
\be
H^{KdV}_{1} = \int \bj \quad ;\quad H^{KdV}_{3} = \int \bj^2
\quad ;\quad H^{KdV}_{5} = \int \( 2 \bj^3 + \bj \bj^{\pr \pr} \)
\lab{kdvham}
\ee
Moreover one checks that the flow equation:
\be
{\d l_J}/ {\d t_r} = \sbr{(l_J^r)_{+} }{l_J}
\lab{flowkdv}
\ee
gives on the lowest level
${\d \bj}/ {\d t_1}= \bj^{\pr}$ and ${\d \bj}/ {\d t_3} =
\bj^{\pr \pr \pr} + 6 \bj \bj^{\pr}$,
with the second equation reproducing the well-known KdV equation.
This equation can also be obtained by inserting $X=\bj$ and $H^{KdV}_{3}$
into \rf{diraham}.
\lskip
{\sl mKdV Hierarchy.}
Now consider the quadratic two-boson hierarchy with Lax given in \rf{sjlax},
\rf{sjlax1} or \rf{sjlax2}.
We choose as a Dirac constraint: $\theta = \sj \,+ \bsj\, = 0$.
The resulting Dirac bracket structure is:
\be
\{ \sj \, (x) \, , \, \sj \, (y) \}_2^{D} =
- \int dz dz^{\pr} \{ \sj \, (x) \, , \, \theta (z) \}_2 \{ \theta (z) , \theta
 (z^{\pr} )
\}_2^{-1} \{ \theta (z^{\pr}) \, , \, \sj \, (y) \}_2 = - \h \d^{\pr} (x-y)
\lab{dira3}
\ee
and the reduced Lax operator is:
\be
l_{\sj} = D - \sj \, D^{-1} \sj \, = D + \sumi{n=0} (-1)^{n+1} \sj\,
\sj^{\,(n)} D^{-1-n}
 \lab{mkdvlax}
\ee
Note that imposing the constraint $\theta=0$ on the equivalent Lax
operators from \rf{sjlax1} and \rf{sjlax2} respectively, we get:
\br
l_{\sj} \eq L_{\sj} \bv_{\theta=0} = D + \sj \(D \,+\,
{\sj^{\,\pr} \o \sj} \)^{-1}  \, \( - \sj\, - \,{\sj^{\, \pr} \o \sj} \)
= D + D^{-1} \( - \sj^{\, 2} - \, \sj^{\, \pr}\)
\lab{sjlaxreda} \\
l_{\sj} \eq L_{\sj} \bv_{\theta=0} = D -\( \sj\, +\,{\sj^{\, \pr} \o \sj} \)
\(D \,-\, {\sj^{\,\pr} \o \sj} \)^{-1}  \, \sj \,
= D +  \( - \sj^{\, 2} - \, \sj^{\, \pr}\) \, D^{-1}
\lab{sjlaxredb}
\er
Obviously we could have expressed everywhere $\sj\,$ by $- \bsj\,$ hence
the one-boson system must be invariant under transformation
$\sj \, \leftrightarrow - \sj \,$.
The flow equations calculated as in \rf{flowkdv} are
\be
{d \sj\, \o d t_1} = \sj\,^{\pr} \quad;\quad {d \sj\, \o d t_2} = 0
\quad ;\quad
{d \sj\, \o d t_3} = \sj\,^{\pr \pr \pr}  + 6 \sj\,^2(\sj\,)^{\pr}
\lab{1bosflow}
\ee
Hence the flow equation for $ d \sj / d t_3$ is the mKdV equation.
Furthermore the mKdV equation could also be obtained from Hamiltonian
$H^{mKdV}_3$ defined in a standard way:
\be
H_{1}^{mKdV} =- \int \sj\,^2 \;\, ;\;\, H_{3}^{mKdV} = \int \(\sj\,^4 -
\sj\,\sj^{\, \pr\pr}\)\;\, ;\;\,H_{5}^{mKdV} = - \int \( 2 \sj\,^6 + 10
\sj\,^2 (\sj\,^{\pr})^2 + \sj\, \sj\,^{(IV)} \)
\lab{mkdvham}
\ee
(and zero for even indices).
Because of existence of symmetry described in \rf{sjtransf} (and
\rf{sjtransf2}) we could equivalently impose the constraints
$\theta_1 = \sj \,+ \bsj\,- \sj^{\, \pr} /\sj\,  = 0$ or alternatively
$\theta_2 = \sj \,+ \bsj\,+\bsj^{\, \pr} /\bsj\, = 0$
without changing the Dirac bracket structure and the constraint manifold.
Imposing $\theta_1= 0 $ on the Lax operator in \rf{sjlax} we get
\be
l_{\sj} = D + \( - \sj\, + {\sj^{\, \pr} \o \sj} \)
\(D \,-\, {\sj^{\,\pr} \o \sj} \)^{-1}  \, \sj
= D + \( - \sj^{\, 2} + \sj^{\, \pr}\) D^{-1}
\lab{sjlaxred1}
\ee
Taking however the equivalent Lax operator as given in \rf{sjlax1}
we get $l_{\sj} = L_{\sj} \bv_{\theta_1 =0} = D - \sj \, D^{-1} \sj\,$.
Hence the mKdV hierarchy is given in terms of three alternative and equivalent
Lax operators given in \rf{mkdvlax}, \rf{sjlaxreda} and \rf{sjlaxred1}.
Especially the mKdV Hamiltonians (including those in \rf{mkdvham}) are
invariant under transformation $ \sj \, \to - \sj \,$.
\lskip
{\sl Miura Map.}
Let us now impose the Dirac
constraint $J = - \sj \,- \bsj \, + \sj^{\,\pr} /\sj \, =0 $ on
the generalized Miura transformation \rf{gmiura}.
As a result we get the conventional Miura map:
\be
\bj\bv_{J=0} = \sj \, \( - \sj \, + {\sj^{\,\pr}  \o \sj} \) =
-\sj\,^2 + \sj^{\,\pr}
\lab{miura}
\ee
It is easy to find via Dirac procedure that $\sj\,$ satisfies the bracket
\be
\{ \sj \, (x) \, , \, \sj \, (y) \}_2^{D} =
- \int dz dz^{\pr} \{ \sj \, (x) \, , \, J (z) \}_2 \{ J (z) , J (z^{\pr} )
 \}_2^{-1}
\{ J (z^{\pr}) \, , \, \sj \, (y) \}_2 = - \h \d^{\pr} (x-y)
\lab{mdira3}
\ee
which is perfectly consistent with $ \bj =-\sj\,^2 + \sj^{\,\pr}$ satisfying
the bracket \rf{dira2}.

Especially we see that all Hamiltonians from \rf{kdvham} go to Hamiltonians
in \rf{mkdvham} under $\bj \to - \sj\,^2 \pm \sj^{\,\pr}$.
\lskip
{\sl Bi-Hamiltonian Structure of KdV Hierarchy.}
The evolution equation \rf{diraham} specified to the constrained manifold
$ J=0$ results in
\be
\partder{\bj}{t_r} \bv_{J=0} = {\pbr{\bj}{H^{KdV}_r}}^D_2 =
\( D \bj + \bj D + \h D^3 \) \funcder{H_r}{\bj}\bv_{J=0}
\lab{kdvbra2}
\ee
in which one recognizes the second Hamiltonian structure of KdV hierarchy.
We now show how to recover the first Hamiltonian structure of the KdV
hierarchy (our discussion is here parallel to that given in \ct{BAK85}).
Recall now \rf{iflow} and take $r$ odd so $H_{r \pm 1} \to 0$ for $J \to 0$.
We find from \rf{iflow} using $P_{J\,1}$ that
$\pa {\bj}/ \pa {t_r} \bv_{J=0} = - D \d {H_{r+1}} /\d {J} \bv_{J=0}$.
On the other hand calculating $\pa {J}/ \pa {t_{r+1}} $
using both $P_{J\,1}$ and $P_{J\,2}$ we find the following consistency
relation in the two-boson case:
\be
2 D \funcder {H_{r+1}}{J} + \( D^2 + D J \)
\funcder {H_{r+1}}{\bj} = - D \funcder{H_{r+2}}{\bj}
\lab{jtr1}
\ee
However in the limit $J\to 0$ since $H_{r+1} \to 0$ we have
$\d H_{r+1}/ \d \bj \to 0$.
Therefore summarizing we find
\be
\partder{\bj}{t_r} \bv_{J=0} = - D \funcder{H_{r+1}}{J} \bv_{J=0}
= \h D \funcder{H^{KdV}_{r+2}}{\bj}
= \(\, \h D^3 + \bj D + D \bj\, \) \funcder{H^{KdV}_r}{\bj} \lab{bihamkdv}
\ee
which reproduces well-known result about bi-Hamiltonian structure of KdV
(see also \ct{BAK85}).
Equation \rf{bihamkdv} can also be treated as a recurrence
relation which proves that the system defined by Lax given in \rf{faa3}
is indeed KdV system to all orders of the Hamiltonian function.

One can now find the bi-Hamiltonian structure for the case of mKdV.
First we recall a formula \ct{wilson}:
\be
\( D \mp 2 \sj \, \) D \( D \pm 2 \sj \, \)= 2 \( \h D^3 + ( -\sj^{\,2} \pm
\sj^{\,\pr}) D + D ( -\sj^{\,2} \pm \sj^{\,\pr}) \)
\lab{wilson}
\ee
Next from Miura transformation we find \ct{wilson}
$\, \d {H^{mKdV}_{r}}/ \d {\sj}\, =  \, \( -D - 2 \sj \, \) \d {H^{KdV}_{r}} /
\d {\bj}$.
We therefore have
\be
\funcder{H^{mKdV}_{r+2}}{\sj} = \( -D - 2 \sj \, \)
\funcder{H^{KdV}_{r+2}}{\bj} =
\( D +2 \sj \, \) D^{-1} \( D -2 \sj \, \) D \funcder {H^{mKdV}_{r}}{\sj}
\lab{frecheta}
\ee
where we used both \rf{bihamkdv} and \rf{wilson}.
Relation \rf{frecheta} reveals a bi-Hamiltonian (but non-local) structure
of mKdV hierarchy and can be rewritten in a more simple way as the Lenard
recursion relation:
\be
D \, \funcder {H^{mKdV}_{r+2}}{\sj}\, =
\( D^3 - 4 D \, \sj \,D^{-1} \sj\,D\, \) \, \funcder {H^{mKdV}_{r}}{\sj}
\lab{mkdvbiham}
\ee
\lskip
{\sl Schwarzian KdV Hierarchy.}
Here few remarks are given about Schwarzian KdV (SKdV) hierarchy,
e.g. \ct{weiss}.
We start by discussing invariance of the Miura map
$ \bj = -\sj^{\, 2} + \sj^{\, \pr} =  -(\p^{\pr})^{\, 2} + \p^{\pr \pr}$
where as before $ \p^{\pr} = \sj\,$.
Let $\d$ be some transformation which leaves $\bj$
invariant, then
$\d \(  -(\p^{\pr})^{\, 2} + \p^{\pr \pr} \) = 0 $
or $\d \p^{\pr \pr} = 2 \p^{\pr}\d \p^{\pr}$.
Solution to this takes a simple form
\be
\d \p^{\pr} = \d \sj \, = \eps^{-1} \exp ( 2 \p ) \qquad {\rm or} \qquad
\d \p = {\eps^{0}\o 2} + \eps^{-1} \exp ( 2 \p )
\lab{dp}
\ee
where $\eps^{0}$ and $ \eps^{-1} $ are some arbitrary constants.
Introduce now the function $f$
connected to $\sj$ through the Cole-Hopf type of transformation
$\bsj \, = \p^{\pr} = f^{\pr \pr} /2 f^{\pr}$ or
$ f^{\pr} = \exp (2 \p)$.
We find that \rf{dp} corresponds to $sl_2$ transformation
$ \d f = \eps^{1}+ \eps^{0} f + \eps^{-1} f^2 $
and leaves $ \bj = S (f)/2$ invariant, where $S(f)$ is a Schwarzian.

It is known that the Cole-Hopf transformation
relates the mKdV hierarchy to the SKdV hierarchy with equation
$f_t /f^{\pr} = S(f)$.
Hence we will be interested in one-boson Lax operator
of the form
\be
L= D - \h { f^{\pr \pr} \o f^{\pr}} D^{-1} \h { f^{\pr \pr} \o f^{\pr}}
\lab{schwarzlax}
\ee
There are many ways of promoting this operator to two-boson system.
If we consider a very simple choice
\be
L= D + \h { f^{\pr \pr} \o f^{\pr}} + \sj \,+ { f^{\pr \pr} \o f^{\pr}}
D^{-1} \sj
\lab{schwarzlax1}
\ee
the second bracket structure is $\pbr{\sj\,(x)}{f(y)} =
-2 f^{\pr} (x) D^{-1}_x \d (x-y)$.
Another choice could be
$ L= D + (f^{\pr \pr} / f^{\pr}) + 2 \rho + \( (f^{\pr \pr} /f^{\pr}) +
\rho\) D^{-1} \rho $
leading to \rf{schwarzlax} under constraint $(f^{\pr \pr} /f^{\pr} ) +
2 \rho =0$.
Defining $ \rho = v^{\pr}$ we can now make contact with quadratic KP hierarchy
by defining a map:
$ \bsj \, = v^{\pr} + (f^{\pr \pr} / f^{\pr}) $ and
$\sj \, =  v^{\pr}$.
Of course ambiguity of \rf{sjtransf} allows equally well
a map: $\bsj \, = v^{\pr} + (f^{\pr \pr} / f^{\pr}) -
(v^{\pr \pr} / v^{\pr})$ and $\sj \, =  v^{\pr}$.
The corresponding bracket equivalent to $P_{\sj\,2}$ in
\rf{p1sjp2sj} is non-local and we find easily e.g.
$\pbr{v(x)}{f(y)} = D^{-1}_x f^{\pr} (x) D^{-1}_x \d (x-y)$.

\end{document}